\documentclass[12pt, a4paper]{amsart}

\usepackage{amsmath,amssymb,amsfonts}
\usepackage{bbold}
\usepackage{epic,eepic}
\usepackage{enumerate}

\usepackage{amsthm}
\usepackage[all]{xy}
\usepackage{caption}
\usepackage{stmaryrd} 
\usepackage[dvipdfmx]{graphicx,color} 
\usepackage{tikz}
\usetikzlibrary{trees}

\theoremstyle{plain}
\newtheorem{thm}{Theorem}[section]
\newtheorem{prop}[thm]{Proposition}

\theoremstyle{definition}
\newtheorem{defn}[thm]{Definition}

\newtheorem{Note}[thm]{Note}

\numberwithin{figure}{section}

\numberwithin{table}{section}

\newcommand{\lspace} {
  \vspace{0.8\baselineskip}
}

\newcommand{\abs}[1]{
  \lvert  #1 \rvert
}

\newdir{ >}{{}*!/-5pt/@{>}}

\definecolor{arrowred}{rgb}{0,0,0} 
\newcommand{\newword}[1]{\textbf{\textit{#1}}}
\newcommand{\empword}[1]{\textit{#1}}
\newcommand{\textred}[1]{#1}
\newcommand{\textblue}[1]{#1}

\newcommand{\upin} {
  \mathrel{ \rotatebox[origin=c]{90}{$\in$} }
}

\numberwithin{equation}{section}

\def \JELclassification {1}
\def \AMSclassification {1}
\def \withName          {1}

\begin{document}

\title[Stochastic Jumps based on Belief and Knowledge]{A Framework for Analyzing\\Stochastic Jumps in Finance\\based on Belief and Knowledge}

\if  \withName     1
\thanks{This work was supported by JSPS KAKENHI Grant Number 26330026.}

\author[T. Adachi]{Takanori Adachi}
\address{Department of Mathematical Sciences,
         Ritsumeikan University,
         1-1-1 Nojihigashi, Kusatsu, Shiga, 525-8577 Japan}
\email{taka.adachi@gmail.com}
\fi  

\date{\today
}


\if \AMSclassification 1
\fi 

\if  \JELclassification 1
\makeatletter
\newcommand{\subjclassname@JEL}{JEL Classification}
\makeatother
\fi 

\maketitle

\begin{abstract}
We introduce a formal language $\mathbf{IE}$ that
is a variant of 
the language
$\mathbf{PAL}$
developed in \cite{benthem2011}
by adding 
a belief operator
and a common belief operator,
specializing to stochastic analysis.
A constant symbol in the language denotes a stochastic process 
so that we can represent several financial events as formulae in the language,
which is expected to be clues
of analyzing the moments that some stochastic jumps
such as financial crises  occur
based on knowledge and belief of individuals or 
those
shared within groups of individuals.  
In order to represent beliefs, we use
$\sigma$-complete Boolean algebras as generalized $\sigma$-algebras.
We use the representation for constructing a model in which 
the interpretations of the formulae written in the language $\mathbf{IE}$ reside.
The model also uses some new categories for integrating several components 
appeared in the theory into one.
\end{abstract}

\section{Introduction}
\label{sec:intro}

When predicting the timing of financial credit risk events,
it would be better if we can forecast them with a \empword{structural approach}
rather than just analyzing them in a reduced-form manner
since the structural approach may show us the mechanism how the events happen
as well as their timing.
However, it seems that the structural approach does not work so well so far for forecasting risk events
just like that we cannot forecast earthquakes very well.
But comparing with earthquakes is a fair excuse?

One of the main differences between financial risk events and natural disasters is
that the former are triggered by an aggregation of \empword{human speculation} while the others are independent of it.

\lspace

In order to handle human speculation, 
we already have a theory of \empword{utility functions}
that represent \empword{human preferences}.
But,it is somehow too simple to represent \empword{human beliefs} and their changes
since preferences may be results of accumulated individual beliefs.

If we try
to handle the beliefs one by one, 
we need a technique of processing them and a theory of analyzing them.
We have and will have a better technique of handling \empword{big data} with high speed computer systems these days
that will help for solving the former issue.
So the remaining is the theory.

\lspace

In this paper, 
we will present a framework for developing such a theory
by providing a \empword{language} that is capable of describing financial phenomena 
and a \empword{model} for interpreting it based on 
\empword{measure-theoretic} probability theory
so that we can apply it to many assets in mathematical finance we have developed.

\lspace

Trying to make languages handle knowledge and belief is not new.
Actually, they have been developed in
the theories of epistemic and doxastic logic
as modal operators,
starting with a seminal book
\cite{hintikka1962}.
The theories provide models of the languages,
and some of them are based on probabilistic interpretation.
However, the probability theory used there is not based on measure theory
and do not fit to applications using modern stochastic theory,
such as mathematical finance theory.

\lspace

In this paper,
we extend the language so that it can handle 
stochastic processes and give a model based on measure theoretic probability theory
with a help of category theory.

\lspace

Let us see an example of formula that we can represent in our language.
\begin{equation}
\label{eq:ex1}
(\mathbf{B}_j (X(\nu) \ge p))
  \land
(\mathbf{B}_k (X(\nu) \le p))
\end{equation}
where $X$ is (a name of) a stochastic process representing a value movement of some stock,
$\nu$ is a built-in constant denoting ``now'',
$p$ is a specified price,
and
$j$ and $k$ are agents.
We read this formula as
``\textit{the agent $j$ 
 believes that the current value of the stock is more than or equal to $p$,
while 
the agent $k$ believes that it is less than or equal to $p$}''.
If the formula is true, there may happen a trade between $j$ and $k$,
that is, the agent $j$ may sell some amount of the stock to the agent $k$ at the price
$p$.

In this paper,
we will provide a model with which we can evaluate the formula.
Here is a formula for the evaluation.
\begin{equation}
\label{eq:exItaJump}
i, \omega, t
  \models_{0.05}
(\mathbf{B}_j (X(\nu) \ge p))
  \land
(\mathbf{B}_k (X(\nu) \le p))
\end{equation}
This says that
``\textit{The agent (an observer) $i$ at state $\omega$ and time $t$ evaluates
 (\ref{eq:ex1})
95\% valid}''.

Here is another example.
\begin{equation}
\label{eq:exCredRisk}
i, \omega, t
  \models_{0.1}
\mathbf{CB}_G
 (V(\nu + 1) \ge \ell)
\end{equation}
where $V$ is (a name of) a stochastic process representing some firm's value,
$\ell$ is its liabilities,
and $G$ is a set of agents (of concerned parties).
This says that
``\textit{The agent $i$ at state $\omega$ and time $t$ evaluates
that it is a common belief among $G$
that the firm's value at $t+1$ is more than or equal to $\ell$
with 90\% degree of certainty}''.
When the common belief breaks down, it may be a point of starting a credit risk event.

In both examples, we have clues of analyzing the moments that some jumps occur
based on \textit{belief},
which may be a new perspective of microeconomics or credit risk theory.

\lspace

In Section \ref{sec:langIE},
we will introduce a formal language
 $\mathbf{IE}$
that can 
depict
the situations we are interested in financial markets
including 
(\ref{eq:exItaJump})
and
(\ref{eq:exCredRisk}).
In order to interpret formulae written in the language,
we need to understand what knowledge and belief are in our setting.
Section
\ref{sec:knowBel}
provides
a solution to it
by using 
$\sigma$-complete Boolean algebras.
After introducing few categories as a preparation in
Section
\ref{sec:moreCats},
we give an interpretation of 
$\mathbf{IE}$
in Section
\ref{sec:modelIE}.

\lspace

For those who are not so familiar with Language theory and Category theory,
please consult 
\cite{benthem2011}
and
\cite{maclane1997},
respectively.

\section{The Language $\mathbf{IE}$}
\label{sec:langIE}

Let
\textred{$\mathcal{T}$}
be a time domain with the least time
\textred{0}.
All the discussions in this paper are
on a filtered measurable space
\begin{equation}
\label{eq:omega}
(
\textred{
  \Omega},
\textred{
  \mathcal{G}},
\textred{
  \mathbb{G}}
 = \{
\textred{
      \mathcal{G}_t
}
  \}_{t \in \mathcal{T}}
)
\end{equation}
that satisfies
\textblue{
$
  \mathcal{G}
=
  \bigvee_{t \in \mathcal{T}} \mathcal{G}_t
$}.

\begin{defn}
\label{defn:IEformulae}
Let
$
\textred{
\mathbf{I}
}
$
be a non-empty finite set of 
\empword{individuals}
or
\empword{agents}.

For a 
set
$A$,
an 
\newword{
$\mathbf{IE}$-formula\footnote{
$\mathbf{IE}$ stands for \textit{InEquality}.
}
}
 on $A$
is defined inductively by the following BNF notation\footnote{ Backus-Naur Form }.
\begin{equation*}
\varphi
  ::=
\textred{m_1 \le m_2}
  \mid
\textred{\lnot} \varphi
  \mid
\varphi \textred{\land} \psi
  \mid
\textred{\mathbf{K}_i} \varphi
  \mid
\textred{\mathbf{CK}_G} \varphi
  \mid
\textred{\mathbf{B}_i} \varphi
  \mid
\textred{\mathbf{CB}_G} \varphi
\end{equation*}
where
$m_1$
and
$m_2$
are
$\textblue{terms}$\footnote{
Terms will be defined in
Definition \ref{defn:langTerms}.
},
$
i \in 
\mathbf{I}
$,
$
G \subset
\mathbf{I}
$,
and
$\varphi$
and
$\psi$
are 
$\mathbf{IE}$-formulae.
The formula
$
m_1 \le m_2
$
is called a \newword{primitive formula},
$
\textred{
\mathbf{IE}(A)
}
$
is the set of all 
$\mathbf{IE}$-formulae
on $A$.
\end{defn}

Now we explain 
the intended meaning of the four modal operators
$\mathbf{K}_i$,
$\mathbf{CK}_G$,
$\mathbf{B}_i$
and
$\mathbf{CB}_G$
one by one.
\lspace

$
\textred{\mathbf{K}_i} \varphi
$
($i$ \empword{knows} that $\varphi$),

$
\textred{\mathbf{CK}_G} \varphi
$
($\varphi$ is a \empword{common knowledge} in the group $G$),

$
\textred{\mathbf{B}_i} \varphi
$
($i$ \empword{believes} that $\varphi$),

$
\textred{\mathbf{CB}_G} \varphi
$
($\varphi$ is a \empword{common belief} in the group $G$).

\lspace

Before going into the detail of the theory,
we will provide a crux of the language theory.
The important point here is to distinguish 
\textit{syntax}
and 
\textit{semantics}
clearly.
Syntax is about a mere sequence of symbols
that does not say anything about its meaning,
while 
semantics is about a real world in which
the sequence determined by the syntax is interpreted.

Here is a figure explaining the situation.
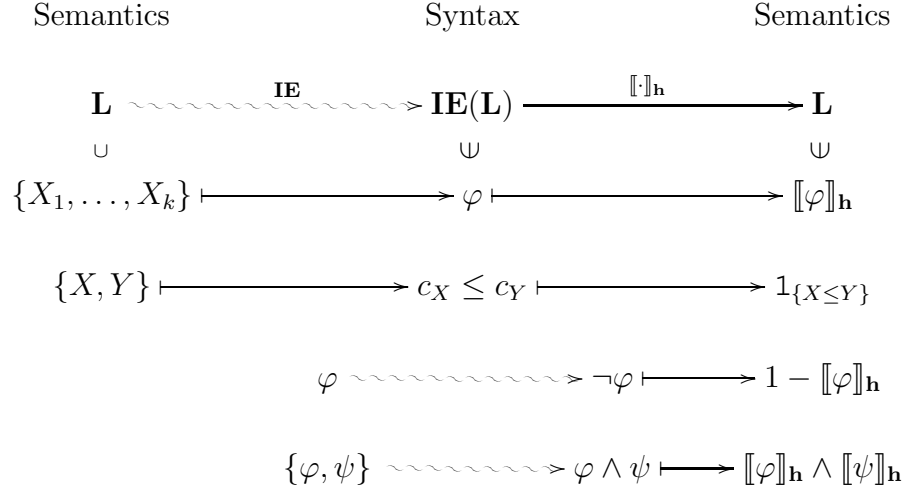
\begin{equation*}
\xymatrix@C=10 pt@R=15 pt{
  \textblue{\textrm{Semantics}}
&&&
  \textred{\textrm{Syntax}}
&&&
  \textred{\textrm{Semantics}}
\\
  \mathbf{L}
     \ar @{~>}^{
        \mathbf{IE}
     } [rrr]
&&&
  \mathbf{IE}(\mathbf{L})
     \ar @{->}^{
        \llbracket \cdot \rrbracket_{\mathbf{h}}
     } [rrr]
&&&
  \mathbf{L}
\\
  \{X_1, \dots, X_k\}
     \ar @{|->} [rrr]
     \ar @{}^{\cup} @<-6pt> [u]
&&&
  \varphi
     \ar @{|->} [rrr]
     \ar @{}^{\upin} @<-6pt> [u]
&&&
  \llbracket \varphi \rrbracket_{\mathbf{h}}
     \ar @{}^{\upin} @<-6pt> [u]
\\
  \{X, Y\}
     \ar @{|->} [rrr]
&&&
  c_X \le c_Y
     \ar @{|->} [rrr]
&&&
  \mathbb{1}_{
    \{X \le Y\}
  }
\\
&&
  \varphi
     \ar @{~>} [rr]
&&
  \lnot \varphi
     \ar @{|->} [rr]
&&
  1 - \llbracket \varphi \rrbracket_{\mathbf{h}}
\\
&&
  \{ \varphi, \psi \}
     \ar @{~>} [rr]
&&
  \varphi \land \psi
     \ar @{|->} [rr]
&&
  \llbracket \varphi \rrbracket_{\mathbf{h}}
    \land
  \llbracket \psi \rrbracket_{\mathbf{h}}
}
\end{equation*}

For example,
a stochastic process 
$X$
is mapped to a constant symbol 
$c_X$
that is a \textit{name} of $X$,
and the sequence of symbols 
$\lnot \phi$
is interpreted as the value
$
  1 - \llbracket \varphi \rrbracket_{\mathbf{h}}
$
using the (already computed) interpretation 
$
  \llbracket \varphi \rrbracket_{\mathbf{h}}
$
of
$\varphi$.

\lspace

First of all,
as a component of the language 
$\mathbf{IE}$,
we see the syntax and the semantics of 
\textit{terms}
appeared in Definition
\ref{defn:IEformulae}.

Here is its syntax.
\begin{defn}
\label{defn:langTerms}
Let 
$X$ be a 
$\mathbb{G}$-adapted process,
$r \in \mathbb{R}$,
and
$f : \mathbb{R}^k \to \mathbb{R}$
be a predefined measurable function.
Then, a 
\newword{term}
$m$
is defined inductively by:
\begin{equation*}
\textred{
m 
}
  ::=
c_r
  \mid
\textred{
\nu
}
  \mid
\textred{c_X}(m)
  \mid
\textred{c_f}(m_1, \dots, m_k)
\end{equation*}

\end{defn}

Then, its semantics is provided as following.
\begin{defn}

The values of a term $m$ is a \empword{stochastic process}
\textred{
$ \llbracket m \rrbracket$
}
 defined by
\begin{enumerate}
\item
$
\textred{
\llbracket
   c_r
\rrbracket}(t) :=
 r
$,

\item
$
\textred{
\llbracket
  \nu
\rrbracket}(t) :=
  t
$,

\item
$
\textred{
\llbracket
  c_X(m)
\rrbracket}(t) :=
  X(\llbracket m \rrbracket(t) )
$,

\item
$
\textred{
\llbracket
  c_f (m_1, \dots, m_k)
\rrbracket}(t) :=
  f (
    \llbracket m_1 \rrbracket(t),
     \dots,
    \llbracket m_k \rrbracket(t)
  )
$.

\end{enumerate}

\end{defn}

\lspace

In order to give semantics for all 
$\mathbf{IE}$-formulae,
we need some considerations about
knowledge and belief
as well as
domains of their models.

\lspace
We sometimes write simply
$\textblue{a}$
for
$\textblue{c_a}$.
We allow the following abbreviations as 
\empword{syntactic sugar}
for
$\mathbf{IE}$-formulae.
\begin{align*}
m_1 \textred{=} m_2
  \; &\equiv \;
(m_1 \le m_2)
  \land
(m_2 \le m_1)
,
\\
m_1 \textred{\ne} m_2
  \; &\equiv \;
\lnot
(m_1 {=} m_2)
,
\\
m_1 \textred{<} m_2
  \; &\equiv \;
(
m_1 \le m_2
)
\land
(m_1 {\ne} m_2)
,
\\
m_1 \textred{\ge} m_2
  \; &\equiv \;
\lnot
(
m_1 < m_2
)
,
\\
m_1 \textred{>} m_2
  \; &\equiv \;
\lnot
(
m_1 \le m_2
)
,
\\
\varphi \textred{\lor} \psi
  \; &\equiv \;
\lnot (
 (\lnot \varphi) \land (\lnot \psi)
)
,
\\
\varphi \textred{\to} \psi
  \; &\equiv \;
(\lnot \varphi) \lor  \psi
,
\\
\varphi \textred{\leftrightarrow} \psi
  \; &\equiv \;
(\varphi \to \psi)
   \land
(\psi \to \varphi)
.
\end{align*}

The set
$
      \mathbf{IE}(A)
$
is just like a 
\empword{free algebra}
 generated by 
$A$.
In that sense, 
the following functor
$
      \mathbf{IE}
:
      \mathbf{Set}
\to
      \mathbf{Set}
$
can become a 
\empword{monad}.
\begin{equation*}
\xymatrix@C=18 pt@R=15 pt{
    \mathbf{Set}
       \ar @{->}^{
          \textred{
            \mathbf{IE}
          }
       } [rr]
  &&
    \mathbf{Set}
\\
    A
     \ar @{->}_{
       f
     } [dd]
     \ar @{|->} [rr]
  &&
      \mathbf{IE}(A)
     \ar @{->}^{
       \textred{
         \mathbf{IE}(f)
       }
     } [dd]
     \ar @{}^-{\ni} @<-6pt> [r]
  &
    \varphi
     \ar @{|->} [dd]
\\\\
    B
     \ar @{|->} [rr]
  &&
      \mathbf{IE}(B)
     \ar @{}^-{\ni} @<-6pt> [r]
  &
    \mathbf{IE}(f)(\varphi)
}
\end{equation*}
where
$
\textred{
    \mathbf{IE}(f)(\varphi)
}
$
is the $\mathbf{IE}$-formula 
on $B$
that we get by
substituting all occurrences of the form
$c_a$
in
$\varphi$
by 
$c_{f(a)}$.

\section{How to Represent Knowledge and Belief}
\label{sec:knowBel}

\subsection{Knowledge and Belief}
\label{sec:subKB}

In this subsection,
we require an extra assumption that
the cardinality of 
$\Omega$
is finite.
But, this requirement is just for explaining our motivation described below,
and is not necessary in general.

\lspace

Let
$\textred{\mathcal{F}} \subset \mathcal{G}$
be a sub-$\sigma$-algebra of $\mathcal{G}$.

\lspace

Since $\Omega$ is a finite set,
a $\sigma$-algebra
$
    \mathcal{F}
$
defines a partition of $\Omega$
whose equivalence classes are of the form
\begin{equation*}
    [\omega]_{\mathcal{F}}
    :=
    \bigcap
    \{
      A \in \mathcal{F}
    \mid
          \omega \in A
    \}
\end{equation*}
from which we can recover the original $\sigma$-algebra.

\lspace

Then, we read this situation as
\lspace

\textit{
``
According to the \textblue{knowledge} $\mathcal{F}$,
we cannot distinguish 
two (fundamental) events
$\omega_1$
and
$\omega_2$
in $\Omega$
if
$
[\omega_1]_{\mathcal{F}}
  =
[\omega_2]_{\mathcal{F}}.
\textrm{''}
$
}
Therefore, if we say that we know something at state $\omega$,
then it means that we recognize that it is true in 
all states $\omega'$  in 
    $[\omega]_{\mathcal{F}}$.
In other words,
the smaller the set
    $[\omega]_{\mathcal{F}}$ is
the more certainly
we know it.

So, 
\textblue{we regard a $\sigma$-algebra as a representation of knowledge}.

\lspace

Now let us go forward to the belief issue.
Plato said,
\textblue{\textit{knowledge $=$ justified true belief}}.
The proverb suggests that
the \textblue{clear-cut} boundary between different equivalence classes
$[\omega_1]_{\mathcal{F}}$
and
$[\omega_2]_{\mathcal{F}}$
specified by the knowledge $\mathcal{F}$
needs to be \textblue{blurry} when making it a representation of \textblue{belief}.

\lspace

Since a subset $A$ of $\Omega$ can be identified with its characteristic function:
\begin{equation}
\label{eq:charFun}
A : \Omega \to \textblue{\{ 0, 1 \}},
\end{equation}
we can think a $\sigma$-algebra as a set of characteristic functions.

Now let us try to generalize 
the range of 
(\ref{eq:charFun})
in order to introduce \textblue{blur}.

A candidate is a $\sigma$-complete Boolean algebra.

\subsection{$\sigma$-complete Boolean algebras}

In this subsection, we will give 
a review of Boolean algebras.
People who want to see more detail,
please refer to
\cite{sikorski1969}.

\begin{defn}
\label{defn:boolAlg}
A 
\newword{Boolean algebra}
is a structure
\begin{equation*}
(B, 0, 1, \land, \lor, \lnot),
\end{equation*}
where 
$
\textred{B}$ is a set,
$
\textred{0}, 
\textred{1} \in B$,
$\textred{\land}$ and 
$\textred{\lor}$
are binary operations on $B$
 and 
$\textred{\lnot}$
is a unary operation on $B$,
 satisfying the following conditions called 
the \newword{Boolean laws}.
\begin{enumerate}
\item 
$
x \land x = x,
\quad
x \lor x = x,
$

\item 
$
x \land y = y \land x,
\quad
x \lor y = y \lor x,
$

\item 
$
x \land (y \land z) = (x \land y) \land z,
\quad
x \lor (y \lor z) = (x \lor y) \lor z,
$

\item 
$
x \land (x \lor y)
=
x \lor (x \land y)
=
x,
$

\item 
$
x \land (y \lor z) = (x \land y) \lor (x \land z),
\quad
x \lor (y \land z) = (x \lor y) \land (x \lor z),
$

\item 
$
x \lor 0 = x,
\quad
x \land 1 = x,
$

\item 
$
x \land 0 = 0,
\quad
x \lor 1 = 1,
$

\item 
$
x \land \lnot x = 0,
\quad
x \lor \lnot x = 1,
$

\item 
$
\lnot \lnot x = x,
$

\item 
$
\lnot(x \land y) = \lnot x \lor \lnot y,
\quad
\lnot(x \lor y) = \lnot x \land \lnot y.
$

\end{enumerate}

\end{defn}

It is easy to show that
for every pair of elements
$
x, y \in B
$
of a Boolean algebra, 
we have
\begin{equation}
x \land y = x
  \; \Leftrightarrow \;
x \lor y = y .
\end{equation}
We write this situation by
$x \textred{\le} y$.
Then, 
the structure
$(B, \le)$
becomes a partilly ordered set.

\begin{defn}
\label{defn:sigmaBoolAlg}
A Boolean algebra
$
(B, 0, 1, \land, \lor, \lnot)
$
is called 
\newword{$\sigma$-complete}
if for all index sets $I$ and $J$ with 
\begin{equation*}
\abs{I},
\abs{J}
  \le
\textblue{\aleph_0},
\end{equation*}
there exist elements of $B$
\begin{equation*}
\textred{\bigwedge_{i \in I}} x_i
  \quad
\textrm{and}
  \quad
\textred{\bigvee_{i \in I}} x_i
\end{equation*}
such that
\begin{equation*}
\bigwedge_{i \in (I \cup J)} x_i
  =
( \bigwedge_{i \in I} x_i )
   \land
( \bigwedge_{i \in J} x_i ) 
\end{equation*}
and
\begin{equation*}
\bigvee_{i \in (I \cup J)} x_i
  =
( \bigvee_{i \in I} x_i )
   \lor
( \bigvee_{i \in J} x_i ) .
\end{equation*}

\end{defn}

We now have some examples of $\sigma$-complete Boolean algebras.
\begin{enumerate}
\item
$
\textred{
\mathbf{2}
}
  =
\{ 0, 1 \}
$
is
a $\sigma$-complete Boolean algebra
with the standard Boolean operation,
where
$0$
and
$1$ 
 stand for \textblue{false}
and \textblue{true},
respectively.

\lspace
\item
For a set
$\Omega$,
Any
$\sigma$-algebra
$\mathcal{G}
  \subset
2^{\Omega}
$
is a $\sigma$-complete Boolean algebra
with the set-operations like for every
$A, B \in \mathcal{G}$,
\begin{align*}
0 &:= \emptyset ,
\\
1 &:= \Omega ,
\\
A \land B &:= A \cap B ,
\\
A \lor B &:= A \cup B ,
\\
\lnot A &:= \Omega - A .
\end{align*}

\item
Let
$
(\Omega, \mathcal{G}, \mu)
$
be a measure space.
Then, the quotient set
\begin{equation*}
\textblue{
\mathcal{G} / \sim_{\mu}
}
\end{equation*}
is a $\sigma$-complete Boolean algebra,
where 
$
\textred{
 \sim_{\mu}
}
$
is the equivalence relation on
$
    \mathcal{G}
$
defined by
\begin{equation*}
S
\textred{
  \sim_{ \mu }
}
T
\; \Leftrightarrow \;
  \mu(
S
  \triangle
T
)
= 0
\end{equation*}
for 
$
S, T \in \mathcal{G}
$.

\lspace
\item
Let $\Omega$ be a set,
and $B$ be a 
$\sigma$-complete Boolean algebra.
Then, the function space
\begin{equation*}
\textblue{
  B^{\Omega}
}
\end{equation*}
becomes a 
$\sigma$-complete Boolean algebra
by extending Boolean operations in pointwise manner.

\end{enumerate}

We will use a sub-$\sigma$-complete Boolean algebra
$
  \mathcal{F}
\subset
  B^{\Omega}
$
as a \textblue{generalized} or \textblue{blurred}  $\sigma$-algebra later.

\begin{defn}
\label{defn:catSigmaBoolAlg}
The category
$
\textred{\sigma\mathbf{BA}}
$
is defined by:
\begin{enumerate}
  \item
    \textbf{objects} $ := $
    all 
    non-degenerate\footnote{
A Boolean algebra
is called \newword{degenerate}
if
$0 = 1$.
}
    $\sigma$-complete Boolean algebras,

  \item
    $
  \textred{
    \sigma\mathbf{BA}
    (A, B)
  }
 := 
    $
    all Boolean structure-preserving functions from $A$ to $B$,
    that is,
    the function $f$ satisfying
\begin{enumerate}
\item
$
f(0) = 0
$,
\item
$
f(\bigwedge_{i \in I}x_i ) = \bigwedge_{i \in I} f(x_i) 
$,
\item
$
f(\bigvee_{i \in I}x_i ) = \bigvee_{i \in I} f(x_i) 
$,
\item
$
f(\lnot x) = \lnot f(x)
$.
\end{enumerate}

\end{enumerate}

\end{defn}

\begin{prop}
\label{prop:catBA}
Let 
$b : B \to \mathbf{2}$
be an arrow in 
$\sigma\mathbf{BA}$.
Then,
\begin{enumerate}
\item
$b^{-1}(1)$ is a \textblue{ultrafilter} of
 $B$,

\item
$b^{-1}(0)$ is a \textblue{prime ideal} of
 $B$,

\item
$\mathbf{2}$
is the \textblue{initial object}
of
$\sigma\mathbf{BA}$.

That is,
for every object
$B \in \sigma\mathbf{BA}$,
there exists one and only one arrow
$
\textred{!_B} : \mathbf{2} \to B
$.

\end{enumerate}
\end{prop}

\subsection{Anchoring of generalized $\sigma$-algebras}
\label{sec:anchoring}

Now we proceed to introduce an important concept of
\textit{anchoring}.

Let
$\textred{\mathbb{B}}$
 be a 
$\sigma$-complete Boolean algebra,
and
$
\textred{\mathcal{F}}
  \subset
\mathbb{B}^{\Omega}
$
be a subalgebra
(
sub-$\sigma$-complete Boolean algebra
).
We regard this
$\mathcal{F}$
as a 
\textit{generalized}
$\sigma$-algebra.

\begin{defn}{[Anchoring]}
\label{defn:anchoring}
\begin{enumerate}
\item
An \newword{anchor}
or a \newword{belief}
from 
$\mathbf{B}$
is an order-preserving arrow
$
a : \mathbb{B} \to \mathbf{2}
$
such that
$a(0) = 0$
and
$a(1) = 1$.

We write
the set of all anchors from 
$
\mathbf{B}
$
by
$
\textred{
  \mathcal{A}_{\mathbf{B}}
}
$.

\item
For a non-empty subset
$
A
   \subset 
\mathcal{A}_{\mathbf{B}}
$,
\begin{equation*}
\textred{
  \mathcal{F}
/
  A
}
  := \{
  u \in \mathbf{2}^{\Omega}
\mid
  \exists k \in \mathcal{F},
\,
  \forall a \in A
\,
  [ u = a \circ k]
\}.
\end{equation*}

\vspace{-0.5cm}

\begin{equation*}
\xymatrix@C=10 pt@R=15 pt{
  \Omega
    \ar @{>}^{k} [rr]
    \ar @{>}_{u} [rd]
&&
  \mathbb{B}
    \ar @{>}^{a} [ld]
\\
&
  \mathbf{2}
}
\end{equation*}

\item
For 
$
 a \in 
\mathcal{A}_{\mathbf{B}}
$,
$
\textred{
  \mathcal{F}
/  a 
}
  :=
  \mathcal{F}
/ \{ a \}
$.

\end{enumerate}
\end{defn}

\begin{prop}
Let
$
A_1, A_2 \subset 
\mathcal{A}_{\mathbf{B}}
$
be non-empty subsets.
\begin{enumerate}
\lspace

\item
If
$
A_1 \subset A_2
$,
then
$
\mathcal{F} / A_1
  \supset
\mathcal{F} / A_2
$.

\lspace
\item
$
\mathcal{F} / 
  \mathcal{A}_{\mathbf{B}}
= \{
  1_{
    \mathcal{A}_{\mathbf{B}}
  }
  \circ k
\mid
  k \in \mathcal{F}
\, \textrm{and} \,
  \forall \omega \in \Omega [
       k(\omega) = 
         0_{\mathbb{B}}
           \, \textrm{or} \,
         1_{\mathbb{B}}
   ]
\}
$.

\end{enumerate}
\end{prop}

\lspace

For a given (blurred)
information
$\mathcal{F} \subset \mathbb{B}^{\Omega}$,
we see
\begin{equation*}
\textred{
  \mathcal{F} / 
    \mathcal{A}_{\mathbf{B}}
}
\end{equation*}
as a \textblue{representation of knowledge},
while
we see
\begin{equation*}
\textred{
  \mathcal{F} / 
    a
}
\end{equation*}
as a \textblue{representation of belief}
where you may change your mind (or belief)
by choosing
$
a \in
    \mathcal{A}_{\mathbf{B}}
$.

\begin{defn}
The category
$
\textred{\chi_{\mathcal{F}}}
  :=
\textred{\chi_{\mathcal{F}}(\Omega)}
$
is defined by:
\begin{enumerate}
  \item
    \textbf{objects} $ := $
    all triples 
$
\textred{
  (B, \mathcal{F}, a)
}
$
of an object
$\textblue{B}$
 of 
$\sigma\mathbf{BA}$,
a sub-$\sigma$-complete Boolean algebra
$
\textblue{
 \mathcal{F}
}
  \subset
B^{\Omega}
$,
and an anchor
$
\textblue{
a
}
 : B \to \mathbf{2}
$
in
$
\sigma\mathbf{BA}
$,

  \item
    $
  \textred{
    \chi_{\mathcal{F}}
    (
(B_1, \mathcal{F}_1, a_1),
(B_2, \mathcal{F}_2, a_2)
)
  }
 :=
$

$
\{
  u \in
\sigma\mathbf{BA}
(B_1, B_2)
  \mid
u \mathcal{F}_1
 \subset \mathcal{F}_2
\; \textrm{and} \;
  a_1 = a_2 \circ u
\}
    $

where
$
\textred{u \mathcal{F}} := \{
  u \circ f 
\mid
  f \in \mathcal{F}
\}
$.

\end{enumerate}
\begin{equation*}
\xymatrix@C=8 pt@R=6 pt{
&&
  B_1
    \ar @{>}^{\textblue{u}} [dd]
    \ar @{>}^{a_1} [rrd]
\\
  \Omega
    \ar @{>}^{f_1} [rru]
    \ar @{>}_{f_2} [rrd]
&&&&
  \mathbf{2}
\\
&&
  B_2
    \ar @{>}_{a_2} [rru]
}
\end{equation*}
\end{defn}

Note that
    $
    \chi_{\mathcal{F}}
    (
(\mathbf{2}, \mathcal{F}_1, !_{\mathbf{2}}),
(\mathbf{2}, \mathcal{F}_2, !_{\mathbf{2}})
)
    $
has at most one element, 
and that
    $
    \chi_{\mathcal{F}}
    (
(\mathbf{2}, \mathcal{F}_1, !_{\mathbf{2}}),
(\mathbf{2}, \mathcal{F}_2, !_{\mathbf{2}})
)
    \ne \emptyset
    $
iff
$
\mathcal{F}_1
  \subset
\mathcal{F}_2
$.

\lspace

In the definition of arrows in 
$\chi_{\mathcal{F}}$,
We use one arrow
$u : B_1 \to B_2$
in order to make restrictions for both
$
 \textblue{u} \mathcal{F}_1
\subset
  \mathcal{F}_2
$
and
$
  a_1 = a_2 \circ \textblue{u}
$.
But the second equation means that
$a_1$
is determined by
$a_2$
once 
$u$
is specified according to the first equation,
which may be too restrictive.

Here is another arrow definition of 
$\chi_{\mathcal{F}}$
for removing the restriction:

\lspace

$
  \textred{
    \chi_{\mathcal{F}}
    (
(B_1, \mathcal{F}_1, a_1),
(B_2, \mathcal{F}_2, a_2)
)
  }
 :=
\{
$
// Non-adopted version

$
\quad
  (\textblue{u}, \textblue{v})
\in
\sigma\mathbf{BA}(B_1, B_2)
  \times
\sigma\mathbf{BA}(B_1, B_2)
  \mid
$

$
\quad
\textblue{u} \mathcal{F}_1
 \subset \mathcal{F}_2
\; \textrm{and} \;
  a_1 = a_2 \circ \textblue{v}
$

$
\}
$.

\vspace{-0.5cm}

\begin{equation*}
\xymatrix@C=8 pt@R=6 pt{
&&
  B_1
    \ar @{>}^{\textblue{u}} [dd]
&
  B_1
    \ar @{>}_{\textblue{v}} [dd]
    \ar @{>}^{a_1} [rrd]
\\
  \Omega
    \ar @{>}^{f_1} [rru]
    \ar @{>}_{f_2} [rrd]
&&&&&
  \mathbf{2}
\\
&&
  B_2
&
  B_2
    \ar @{>}_{a_2} [rru]
}
\end{equation*}

However, this version of 
$
\chi_{\mathcal{F}}
$
will fail to make the following correspondence
$
\mathcal{B}
$
a functor,
so we do not adopt this extended version.

\begin{defn}{[Category $\sigma\mathbf{Alg}_{\Omega}$]}
The category
$
\textred{
\sigma\mathbf{Alg}_{\Omega}
}
$
is defined by:
\begin{enumerate}
  \item
    \textbf{objects} $ := $
    all $\sigma$-algebra over $\Omega$ 

  \item
$
\textred{
\sigma\mathbf{Alg}_{\Omega}
(
\mathcal{F}_1,
\mathcal{F}_2
)
}
$
has exactly one arrow if
$
\mathcal{F}_1
  \subset
\mathcal{F}_2
$,
otherwise empty.

\end{enumerate}

\end{defn}

\begin{defn}{[Knowledge Functor $\mathcal{K}$]}
The functor
$\textred{\mathcal{K}}$
is defined by the following diagram.
\vspace{-0.2cm}
\begin{equation*}
\xymatrix@C=18 pt@R=15 pt{
  &&
    \chi_{\mathcal{F}}
       \ar @{->}^{
          \textred{
            \mathcal{K}
          }
       } [rr]
  &&
    \sigma\mathbf{Alg}_{\Omega}
\\
  &
    B_1
     \ar @{->}_{u} [dd]
  &
    (B_1, \mathcal{F}_1, a_1)
     \ar @{->}_{
       u
     } [dd]
     \ar @{|->} [rr]
  &&
    \textblue{
      \mathcal{F}_1 / \mathcal{A}_{B_1}
    }
     \ar @{->}^{
       \textred{
         \mathcal{K}(u)
       }
     } [dd]
\\
    \Omega
      \ar @{->}^{f_1} [ru]
      \ar @{->}_{f_2} [rd]
\\
  &
    B_2
  &
    (B_2, \mathcal{F}_2, a_2)
     \ar @{|->} [rr]
  &&
    \textblue{
      \mathcal{F}_2 / \mathcal{A}_{B_2}
    }
}
\end{equation*}

\end{defn}

\begin{defn}{[Belief Functor $\mathcal{B}$]}
The functor
$\textred{\mathcal{B}}$
is defined by the following diagram.
\begin{equation*}
\xymatrix@C=18 pt@R=15 pt{
  &&&
    \chi_{\mathcal{F}}
       \ar @{->}^{
          \textred{
            \mathcal{B}
          }
       } [rr]
  &&
    \sigma\mathbf{Alg}_{\Omega}
\\
  &
    B_1
     \ar @{->}_{u} [dd]
     \ar @{->}^{a_1} [rd]
  &&
    (B_1, \mathcal{F}_1, a_1)
     \ar @{->}_{
       u
     } [dd]
     \ar @{|->} [rr]
  &&
    \textblue{
      \mathcal{F}_1 / a_1
    }
     \ar @{->}^{
       \textred{
         \mathcal{B}(u)
       }
     } [dd]
\\
    \Omega
      \ar @{->}^{f_1} [ru]
      \ar @{->}_{f_2} [rd]
  &&
    \mathbf{2}
\\
  &
    B_2
     \ar @{->}_{a_2} [ru]
  &&
    (B_2, \mathcal{F}_2, a_2)
     \ar @{|->} [rr]
  &&
    \textblue{
      \mathcal{F}_2 / a_2
    }
}
\end{equation*}

\end{defn}

Note that there is a natural transformation
\begin{equation*}
  \mathcal{K} \dot{\to} \mathcal{B}
\end{equation*}
that represents natural inclusion arrows.

\section{More Categories}
\label{sec:moreCats}

The following categories were introduced in
\cite{adachi_2014crm}
though the category
$
\chi_{\mathcal{F}}(\Omega)
$
in this paper is an extended version.

\begin{defn}{[Categories $\chi_{\mathbb{P}}$ and $\chi$]}
\label{defn:c_chi2}
\begin{enumerate}
\item
$
  \textred{
    \chi_{\mathbb{P}}
  }
    := 
  \textred{
    \chi_{\mathbb{P}}
    ( \mathcal{G} )
  }
$
  \begin{enumerate}
  \item
    \textbf{objects} $ := $
    the set of all probability measures defined on 
    $(\Omega, \mathcal{G})$,
    
  \item
    $
  \textred{
    \chi_{\mathbb{P}}
    (\mu, \nu)
  }
    $
  has exactly one arrow 
  if
    $ \mu \gg \nu $,
  otherwise empty,
  \end{enumerate}
where
$
       \mu
     \textred{
       \gg
     }
       \nu
$
means that
$
       \nu
$
is absolutely continuous to
$
       \mu
$.

\item
$
\textred{
  \chi
}
    := 
\textred{
  \chi(\Omega, \mathcal{G})
}
    := 
  \chi_{\mathcal{F}}( \Omega )
     \times
  \chi_{\mathbb{P}}( \mathcal{G} ).
$

For an object
$
  \mathcal{U}
    \in 
  \chi
$,
we write
\begin{equation*}
 \mathcal{U}
    =:
   ( 
    (
     \textred{\mathbb{B}_{\mathcal{U}}}, 
     \textred{\mathcal{F}_{\mathcal{U}}}, 
     \textred{a_{\mathcal{U}}}
    ), 
     \textred{\mathbb{P}_{\mathcal{U}}}
   ).
\end{equation*}

\item
For an object
$
  \mathcal{U}
    \in 
  \chi
$,
\begin{equation*}
  \textred{
    \mathcal{K}_{\mathcal{U}} 
  }
:=
  \mathcal{K}(p_1(\mathcal{U})),
\quad
  \textred{
    \mathcal{B}_{\mathcal{U}} 
  }
:=
  \mathcal{B}(p_1(\mathcal{U})).
\end{equation*}
where
$
p_1
  : \chi \to \chi_{\mathcal{F}}
$
is a projection functor.

\end{enumerate}
\end{defn}

\begin{defn}{[Functor $\mathbb{L}_{\mathcal{F}}$]}
\label{defn:funLF}
The functor
\textred{
$\mathbb{L}_{\mathcal{F}}$
}
is defined by the following diagram.
\begin{equation*}
\xymatrix@C=18 pt@R=10 pt{
    \sigma\mathbf{Alg}_{\Omega}
       \ar @{->}^{
          \textred{
            \mathbb{L}_{\mathcal{F}}
          }
       } [rr]
  &&
    \mathbf{Set}
\\
    \mathcal{F}_1
     \ar @{->}_{*} [dd]
     \ar @{|->} [rr]
  &&
    \textred{
      \mathbb{L}_{\mathcal{F}}
        (\mathcal{F}_1)
    }
     \ar @{}^-{:=} @<-6pt> [r]
     \ar @{->}^{
       \textred{
         \mathbb{L}_{\mathcal{F}}(*)
       }
     } [dd]
  &
    L^{\infty}( 
      \Omega,
      \mathcal{F}_1
    )
\\\\
    \mathcal{F}_2
     \ar @{|->} [rr]
  &&
    \textred{
      \mathbb{L}_{\mathcal{F}}
        (\mathcal{F}_2)
    }
     \ar @{}^-{:=} @<-6pt> [r]
  &
    L^{\infty}( 
      \Omega,
      \mathcal{F}_2
    )
}
\end{equation*}

\end{defn}

\begin{defn}
We regard the set
\textred{
$\mathcal{T}$
}
with the natural total order
as a category.
\end{defn}

\begin{defn}
An object of the functor category
$\chi_{\mathcal{F}}^{\mathcal{T}}$
is called a
\newword{filtration}.
\end{defn}

\begin{defn}

The 
functor
$
(\mathbb{L}_{\mathcal{F}} \circ \mathcal{B})^{\mathcal{T}}
$
makes the following diagram commute.
\begin{equation*}
\xymatrix@C=15 pt@R=3 pt{
    \chi_{\mathcal{F}}^{\mathcal{T}}
       \ar @{->}^{
          \textred{
            (\mathbb{L}_{\mathcal{F}} \circ \mathcal{B})^{\mathcal{T}}
          }
       } [rrr]
  &&&
    \mathbf{Set}^{\mathcal{T}}
\\
    f
     \ar @{|->} [rrr]
     \ar @{->} [dd]
  &&&
    \textred{
      \mathbb{L}_{\mathcal{F}} \circ \mathcal{B} \circ f
    }
     \ar @{}^-{:=} @<-6pt> [r]
     \ar @{->} [dd]
  &
    \prod_{
       t \in \mathcal{T}
    }
      \mathbb{L}_{\mathcal{F}}(\mathcal{B}(f(t)))
\\\\
    g
     \ar @{|->} [rrr]
  &&&
    \textred{
      \mathbb{L}_{\mathcal{F}} \circ \mathcal{B} \circ g
    }
     \ar @{}^-{:=} @<-6pt> [r]
  &
    \prod_{
       t \in \mathcal{T}
    }
      \mathbb{L}_{\mathcal{F}}(\mathcal{B}(g(t)))
}
\end{equation*}
An element of 
$
      \mathbb{L}_{\mathcal{F}} \circ \mathcal{B} \circ f
$
is called an 
\newword{$f$-adapted process}.

\end{defn}

\section{An Interpretation of $\mathbf{IE}$}
\label{sec:modelIE}

\begin{defn}{[Category $\mathbf{I}$]}
\label{defn:catI}
\begin{enumerate}
\item
We regard the set
\textred{
$\mathbf{I}$
}
of agents 
as a \textblue{discrete category}, 
that is, a category whose objects are elements of the set
$\mathbf{I}$,
but it has no arrows except identities.

\lspace
\item
The function 
$
\textred{\rho} : 2^{\mathbf{I}} \to [0, 1]
$
is a probability measure on
$
(\mathbf{I}, 2^{\mathbf{I}})
$.

\end{enumerate}
\end{defn}

\begin{defn}{[Histories]}
A \newword{history}
is a functor
from $\mathbf{I}$
to
$
  \chi^{\mathcal{T}}
$.

\end{defn}

\lspace

Let
$
\textblue{
  \mathbf{h}
}
  :
\mathbf{I}
  \to
\chi^{\mathcal{T}}
$
be a history.
Then, for
$
i \in \mathbf{I}
$
and
$
t \in \mathcal{T}
$,
we have
an object of $\chi$ like
\begin{equation*}
 \mathbf{h}(i)(t)
    =
   ( 
    (
     \textblue{\mathbb{B}_{\mathbf{h}(i)(t) }}, 
     \textblue{\mathcal{F}_{\mathbf{h}(i)(t)}},
     \textblue{a_{ \mathbf{h}(i)(t)}}
    ),
     \textblue{\mathbb{P}_{\mathbf{h}(i)(t)}}
   ).
\end{equation*}

\lspace

\begin{defn}
Let
$
\textblue{
 \mathbf{h}
}
  :
\mathbf{I}
  \to
\chi^{\mathcal{T}}
$
be a
 history
\begin{enumerate}
\item
A history 
$\mathbf{h}$
is called
\newword{pre-$\mathbb{G}$-adapted}
if
for every
$i \in \mathbf{I}$
and
$t \in \mathcal{T}$,
\begin{equation*}
\mathcal{K}_{\mathbf{h}(i)(t)}
  \subset
\mathbb{G}(t).
\end{equation*}

\item
A 
pre-$\mathbb{G}$-adapted
history 
$\mathbf{h}$
is called
\newword{$\mathbb{G}$-adapted}
if
for every
$i \in \mathbf{I}$
and
$t \in \mathcal{T}$,
\begin{equation*}
\mathcal{B}_{\mathbf{h}(i)(t)}
  \subset
\mathbb{G}(t).
\end{equation*}

\end{enumerate}
\end{defn}

\begin{defn}{[An Interpretation of $\mathbf{IE}$ formulae]}
\label{defn:model2IE}

Let
$
\textblue{
 \mathbf{h}
}
  :
\mathbf{I}
  \to
\chi^{\mathcal{T}}
$
be a
$\mathbb{G}$-adapted
 history
and
$
\textblue{
\varphi
}
\in
\mathbf{IE}(
  \mathbb{L}_{\mathcal{F}}^{\mathcal{T}}(\mathbb{G})
)
$.
Then, the 
\newword{
interpretation
}
 of 
$\varphi$
with the history
$\mathbf{h}$
is a $\mathbb{G}$-adapted stochastic process
\begin{equation*}
\textred{
    \llbracket \varphi \rrbracket_{\mathbf{h}}
}
\in
  \mathbb{L}_{\mathcal{F}}^{\mathcal{T}}(\mathbb{G})
\end{equation*}
defined recursively 
on the structure of 
$\mathbf{IE}$-formula 
by:
\begin{enumerate}
  \item
  $
  \textred{
  \llbracket 
    m_1 \le m_2
  \rrbracket_{\mathbf{h}}
  }(t)
    :=
    \mathbf{1}_{
    \{
      \llbracket
        m_1
      \rrbracket
      (t)
         \le
      \llbracket
        m_2
      \rrbracket
      (t)
    \}
    }
  $,

  \item
  $
  \textred{
  \llbracket 
    \lnot \varphi
  \rrbracket_{\mathbf{h}}
  }
    :=
    1 - 
  \llbracket 
    \varphi
  \rrbracket_{\mathbf{h}}
  $,
  
  \item
  $
  \textred{
  \llbracket 
    \varphi \land \psi
  \rrbracket_{\mathbf{h}}
  }
    :=
  \llbracket
     \varphi 
  \rrbracket_{\mathbf{h}}
    \land
  \llbracket
     \psi
  \rrbracket_{\mathbf{h}}
  $,

  \item
  $
  \textred{
  \llbracket 
    \textbf{K}_i
    \varphi
  \rrbracket_{\mathbf{h}}
  }
  (t)
    := 
  \mathbb{E}^{
    \mathbb{P}_{\mathbf{h}(i)(t)}
  }\big[
    \llbracket 
      \varphi
    \rrbracket_{\mathbf{h}}
    (t)
  \mid
    \textblue{
      \mathcal{K}_{\mathbf{h}(i)(t)}
    }
  \big]
  $,
  
  \item
  $
  \textred{
  \llbracket 
    \textbf{B}_i
    \varphi
  \rrbracket_{\mathbf{h}}
  }
  (t)
    := 
  \mathbb{E}^{
    \mathbb{P}_{\mathbf{h}(i)(t)}
  }\big[
    \llbracket 
      \varphi
    \rrbracket_{\mathbf{h}}
    (t)
  \mid
    \textblue{
      \mathcal{B}_{\mathbf{h}(i)(t)}
    }
  \big]
  $,

  \item
  $
  \textred{
  \llbracket 
    \textbf{CK}_G
    \varphi
  \rrbracket_{\mathbf{h}}
  }
  $
  is a maximal fixed point $\textblue{f}$ of the equation:
\begin{equation}
\label{eq:mfpK}
\textred{f}(t) =
\mathbb{E}^{\rho}\big[
  \mathbb{E}^{\mathbb{P}_{\mathbf{h}(\cdot)(t)}} \big[
     \textred{f}(t)
       \land
     \llbracket
        \varphi
     \rrbracket_{\mathbf{h}} (t)
  \mid
     \textblue{\mathcal{K}}_{\mathbf{h}(\cdot)(t)}
  \big],
  G
\big]
\end{equation}
for all 
$t \in \mathcal{T}$,
  
  \lspace
  \item
  $
  \textred{
  \llbracket 
    \textbf{CB}_G
    \varphi
  \rrbracket_{\mathbf{h}}
  }
  $
  is a maximal fixed point $\textblue{f}$ of the equation:
\begin{equation}
\label{eq:mfpB}
\textred{f}(t) =
\mathbb{E}^{\rho}\big[
  \mathbb{E}^{\mathbb{P}_{\mathbf{h}(\cdot)(t)}} \big[
     \textred{f}(t)
       \land
     \llbracket
        \varphi
     \rrbracket_{\mathbf{h}} (t)
  \mid
     \textblue{\mathcal{B}}_{\mathbf{h}(\cdot)(t)}
  \big],
  G
\big]
\end{equation}
for all 
$t \in \mathcal{T}$.

\end{enumerate}

\end{defn}

In order to guarantee the existence of 
the fixed point solution of 
(\ref{eq:mfpK}),
let us think the following sequence of processes
$
f_n : \mathcal{T} \times \Omega \to [0, 1]
$
defined by
\begin{enumerate}
\item
$
f_0(t)(\omega) := 1
$,

\item
$
f_{n+1}(t) =
\mathbb{E}^{\rho}\big[
  \mathbb{E}^{\mathbb{P}_{\mathbf{h}(\cdot)(t)}} \big[
     f_n(t)
       \land
     \llbracket
        \varphi
     \rrbracket_{\mathbf{h}} (t)
  \mid
     \textblue{\mathcal{K}}_{\mathbf{h}(\cdot)(t)}
  \big],
  G
\big]
$
for
$
n \in \mathbb{N}
$.
\end{enumerate}
Then,
obviously
$\{f_n\}_{n \in \mathbb{N}}$
is a non-increasing sequence in
$[0, 1]^{\mathcal{T} \times \Omega}$.
So, it has a limit,
which is easily proved to be a maximal fixed point of 
(\ref{eq:mfpK}).

The existence of (\ref{eq:mfpB})
is guaranteed by the same reason.

\begin{defn}{[$\varepsilon$-validity of $\mathbf{IE}$ formulae]}
\label{defn:valIE2form}
Let
$
\varepsilon \in [0, 1]
$.
\begin{enumerate}
\lspace
\item
$
i, \omega, t
  \textred{
    \models_{\varepsilon}
  }
\varphi
     \; \Leftrightarrow \;
\mathbb{E}^{\textblue{\mathbb{P}_{\mathbf{h}(i)(t)}}} \big[
   \llbracket \varphi \rrbracket_{\mathbf{h}}(t)
  \mid
   \textblue{
     \mathcal{K}_{\mathbf{h}(i)(t)}
   }
\big](\omega)
  \ge 1 - \varepsilon
$,

\lspace
\item
$
i, \omega, t
  \textred{
    \models
  }
\varphi
     \; \Leftrightarrow \;
i, \omega, t
  \models_{0}
\varphi
$.

\end{enumerate}
\end{defn}

\begin{Note}
In general, we cannot prove that
\begin{equation}
\label{eq:KimpB}
i, \omega, t
    \models
\mathbf{K}_j \varphi
  \to
\mathbf{B}_j \varphi
\end{equation}
since the implementations of 
$ \mathbf{K}_j \varphi $
  and
$ \mathbf{B}_j \varphi $
are using 
\textblue{conditional expectations}
instead of using
\textblue{$\sup$ operations},
failing to have some monotonicity.

If we really insist on 
(\ref{eq:KimpB}),
we may need to assume that the cardinality of
$\Omega$ is finite,
which is not so interesting
from a measure-theoretic point of view.

\end{Note}

\section{Concluding Remarks}
\label{sec:conclusions}

The version of this paper is obviously a very starting point of 
the trial to utilize \textit{belief} for analyzing stochastic jumps.

We introduced a language and its model.
The language can represent some financial situations like the following
and provided a model to interpret them.
\begin{equation*}
i, \omega, t
  \models_{0.05}
(\mathbf{B}_j (X(\nu) \ge p))
  \land
(\mathbf{B}_k (X(\nu) \le p))
,
\end{equation*}
\begin{equation*}
i, \omega, t
  \models_{0.1}
\mathbf{CB}_G
 (V(\nu + 1) \ge \ell)
.
\end{equation*}
The model of the language can treat full theory of stochastic processes
based on measure theory so that one can apply 
several results in mathematical finance theory we have developed for the further analyses.

\lspace

In the future research, 
we have two directions, practical and theoretical issues.

As the practical issue,
we should write code for computing agents' beliefs
from which we can derive the timing of stochastic jumps in some structural way.
And then,
we investigate more concrete examples aiming applications to finance theory.

As the theoretical issue,
we should make a thorough investigation around 
(\ref{eq:KimpB}).
It is also worth to think a possibility to introduce 
a \textit{public announcement} operator
that is quite popular in Epistemic logic 
and is corresponding to a filtration enlargement.

\nocite{FS2011}
\nocite{maclane1997}
\nocite{gilboa2009}
\nocite{adachi_2014crm}

\providecommand{\bysame}{\leavevmode\hbox to3em{\hrulefill}\thinspace}
\providecommand{\MR}{\relax\ifhmode\unskip\space\fi MR }
\providecommand{\MRhref}[2]{%
  \href{http://www.ams.org/mathscinet-getitem?mr=#1}{#2}
}
\providecommand{\href}[2]{#2}

\end{document}